\documentclass[review]{elsarticle}
\usepackage{graphicx}
\usepackage{epstopdf}

\journal{Physics Review A}

\begin{document}
\title{Probing quantum dissipative chaos using purity}

\begin{frontmatter}
\author{A.~R.~Shahinyan}
\ead{anna\_shahinyan@ysu.am, tel/fax: (+37410)575612}
\address{Yerevan State University, Alex Manoogian 1, 0025,
Yerevan, Armenia}

\author{Lock Yue Chew}
\ead{lockyue@ntu.edu.sg, tel/fax(+65) 6316 6984}
\address{School of Physical and Mathematical Sciences, Nanyang Technological University, 21 Nanyang Link, SPMS-PAP-04-04, Singapore 637371}

\author{G.~Yu.~Kryuchkyan}
\ead{kryuchkyan@ysu.am, tel/fax: (+37410)575612}
\address{Yerevan State University, Alex Manoogian 1, 0025,
Yerevan, Armenia}\address{Institute for Physical Researches,
National Academy of Sciences,\\Ashtarak-2, 0203, Ashtarak,
Armenia}

\begin{abstract}
{In this paper, the purity of quantum states is applied to probe chaotic dissipative dynamics. To achieve this goal, a comparative analysis of regular and chaotic regimes of nonlinear dissipative oscillator (NDO) are performed on the base of excitation number and the purity of oscillatory states. While the chaotic regime is identified in our semiclassical approach by means of strange attractors in Poincar\'e section and with the Lyapunov exponent, the state in the quantum regime is treated via the Wigner function. Specifically, interesting quantum purity effects that accompany the chaotic dynamics are elucidated in this paper for NDO system driven by either: (i) a time-modulated field, or (ii) a sequence of pulses with Gaussian time-dependent envelopes.}
\end{abstract}


\end{frontmatter}

\maketitle

\section{Introduction}
Nonlinear dissipative systems demonstrating chaotic behaviour in their dynamics are still the subject
of much attention \cite{Haake}, \cite{Braun}, \cite{Cvitanovich }. The early studies of dissipative 
chaotic systems date back to the works \cite{Graham}.  The investigations
of quantum chaotic systems are distinctly connected with
the quantum-classical correspondence problem in general,
and with environment induced decoherence and dissipation
in particular. It has been recognized \cite{Zurek} that decoherence
has rather unique properties for systems whose classical analogs display chaos. 
Several methods have been proposed to determine whether a quantum dissipative system exhibits chaotic
behaviour. At this point, we note that quite generally, chaos
in classical conservative and dissipative systems with noise,
has completely different properties. For example, strange attractors on Poincar\'e section
can appear only in dissipative systems. The most successful approach that probes quantum dissipative chaos seems to be quantum tomographic methods based on the measurement of the Wigner function, which is a quantum quasi-distribution in phase-space.  In this way, the connection
between quantum and classical treatment of chaos can be realized by means of a comparison between strange attractors
in the classical Poincar\'e section and the contour plots of
the Wigner functions \cite{PhysE},  \cite{mpop}, \cite{Leon}.  However, such manifestation of chaos seems to be hardly realized by experiments because Wigner
function can only be obtained through data post-processing. On the other hand, alternative methods that probe quantum dissipative chaos involve considerations of entropic characteristics, analysis of statistics of excitation number \cite{K2}, \cite{K3}, methods based on the fidelity decay \cite{Lloyd}, and Kullback-Leibler quantum divergence \cite{Kulback}.

In this paper, we discuss an indirect method that reveals quantum dissipative chaos based on an analysis of the purity of states. Purity is a quantity that measures the statistical characteristic of states and decoherence. It is defined through the density matrix $\rho$ of the system as $P=Tr(\rho^2)$. In consequence, the purity of any pure state is 1 and mixed states less than 1.

The goal of this paper is to investigate the variation of the purity of quantum states in regular and chaotic systems. We shall show that purity allows us to distinguish
between the ordered and chaotic quantum dissipative dynamics. In particular, we demonstrate that this program can be realized through the consideration of two schemes of nonlinear dissipative oscillator (NDO): one driven by a continuously modulated pump field, while the other under the action of a periodic sequence of identical pulses with Gaussian envelopes.

\section{Purity and models of nonlinear oscillators}

The purity of the quantum states, which is defined via the density matrix of the system as $Tr(\rho^2)$, is connected to the linear entropy $S_L$ in the following manner:
\begin{equation}
S_L = 1 - Tr(\rho^2)
\end{equation}
and is related to the linear quantum divergence derived from the Kullback-Leibler quantum divergence  \cite{Kulback}.
Note that $S_L$ can be obtained from the von Neumann entropy 
\begin{equation}
S = Tr(\rho ln(\rho))
\end{equation}
as a lower-order approximation. 

For an ensemble of mixed states the density matrix reads as
\begin{equation}
\rho = \sum P_{\psi_j} |\psi _j\rangle \langle\psi_j|,
\label{dm}
\end{equation}
where $P_{\psi_j}$ is the probability of occurrence of state $\psi_j$. In this case, 
the purity takes the form
\begin{equation}
P=Tr(\rho^2) = \sum_{l=0}^{\infty} P_{\psi_j}^2.
\label{dm_2}
\end{equation}
In particular, for thermal light with a photon population $P_{\psi_j}=\frac{\overline{n}^{n}} {(\overline{n}+1)^{(1+n)}}$, the purity can be calculated as
\begin{equation}
P= \frac{1}{2\overline{n} + 1} ,
\label{purity2}
\end{equation}
where $\overline{n}$ is the mean number of thermal photons.
From Eq. \ref{purity2} it is evident that purity decreases as excitation number increases. 

In this paper, we employ quantum purity as a tool to analyse quantum
dynamics of NDO which allows us to determine whether or not the system has reached the chaotic regimes. It is easy to realize that in general the purity of an ensemble of oscillatory states strongly depends on the operational regimes of NDO. More specifically, an increase in the excitation numbers would raise the number of mixing oscillatory states, leading to a decrease in purity which is apparent from Eq. \ref{purity2}. In addition, diffusion and decoherence of oscillatory modes are also relevant to the level of purity. Thus, in the following, we analyse the purity using the density matrix of NDO $\rho(t)$ for both the regular and chaotic regimes. 

We consider a model of anharmonic oscillator driven by external field with time-modulated amplitude that is based on the following Hamiltonian in the rotating wave approximation:
\begin{equation}
H=\hbar \Delta a^{+}a + \hbar (a^{+}a)^{2}+\hbar \left(f(t)a^{+}+
f(t)^{*}a \right), \label{hamiltonian}
\end{equation}
where $a^{+}$ and $a$ are the oscillatory creation and annihilation operators respectively,
$\chi$ is the nonlinearity strength, and $\Delta=\omega_{0} -\omega$ is the
detuning between the mean frequency of the driving field and the frequency of
the oscillator. In the case of a constant amplitude, i.e. $f(t)=\Omega$, this Hamiltonian describes a nonlinear oscillator driven by a monochromatic force. In this paper, we shall consider two cases of driving force: (i) $f(t)=f_0+f_1\exp(\delta t)$, where $\delta$ is the frequency of modulation; and (ii)
\begin{equation}
f(t)=\Omega\sum{e^{-(t - t_{0} - n\tau)^{2}/T^{2}}}  \label{driving} \,.
\end{equation}
For the latter case, the time dependent interaction term is proportional to
the amplitude of the driving field $\Omega$, and consists of Gaussian pulses with
duration $T$ separated by time intervals $\tau$.

The evolution of the system of interest is governed by
the following master equation for the reduced density matrix of the oscillatory mode
in the interaction picture:

\begin{equation}
\frac{d\rho}{dt} =-\frac{i}{\hbar}[H, \rho] +
\sum_{i=1,2}\left( L_{i}\rho
L_{i}^{+}-\frac{1}{2}L_{i}^{+}L_{i}\rho-\frac{1}{2}\rho L_{i}^{+}
L_{i}\right)\label{master},
\end{equation}
where $L_{1}=\sqrt{(N+1)\gamma}a$ and $L_{2}=\sqrt{N\gamma}a^+$ are the
Lindblad operators, $\gamma$ is a dissipation rate, and $N$ denotes the mean
number of quanta of a heath bath. To study the pure quantum effects, we focus below on cases of very low reservoir
temperatures which, however, still ought to be larger than the characteristic
temperature $\mathcal{T} \gg \mathcal{T}_{cr}=\hbar\gamma/k_B$. 

It should be noted that in this approach dissipation and decoherence effects  are described simultaneously by the same Lindblad terms in the Eq. (8). Thus, this approach allows the control of dissipation and decoherence in NDO under time-modulated driving. Note, that quantum effects in  NDO with a time-modulated driving force have been studied in the context of quantum stochastic resonance \cite{AMK}, quantum dissipative chaos \cite{PhysE}-
\cite{K3}, quantum interference assisted by a bistability \cite{qsch} and generation of superposition of Fock states in the presence of decoherence due to the kicking of oscillatory dynamics \cite{Fock}.  

In the semiclassical approach, the corresponding equation of motion
for the dimensionless mean amplitude of oscillatory mode $ \alpha=Tr(\rho a) $ has the following form:
\begin{equation}
\frac{d\alpha}{dt}= -i[\Delta + \chi + 2|\alpha|^2\chi]\alpha + if(t)\Omega -\gamma\alpha.\label{semclass}
\end{equation}
This equation modifies the standard Duffing equation in the case of NDO with time-dependent coefficient. 

These models seem experimentally feasible and can be realized in
several experimental schemes. We mention  nano-electromechanical systems and nano-optomechanical systems based on various nonlinear oscillators \cite{craig}, \cite{ekinci}, and superconducting devices based on the nonlinearity of the
Josephson junction (JJ) \cite{claud}, \cite{vijay}, \cite{Nori} that exhibits a wide variety of quantum oscillatory phenomena.

\section{Purity as an indicator of chaos} 

In this section, we investigate the purity of quantum oscillatory states for various regimes of NDO by performing detailed comparative analysis of purity for both  cases of regular and chaotic dynamics. In general, the purity of an ensemble of oscillatory states strongly depends on the level of the excitation numbers which is particularly apparent from Eq. (\ref{purity2}). Therefore, in our comparative analysis, we consider regimes of regular and chaotic dynamics with the same levels of oscillatory excitation numbers. We shall consider two schemes of nonlinear dissipative oscillator (NDO) in this analysis: a NDO driven by a continuously modulated pump field; and a NDO under the action of a periodic sequence of identical pulses with Gaussian envelopes.

The time evolution of NDO driven by an external time-modulated force  cannot be solved analytically, and hence suitable numerical methods have to be employed. Indeed, only NDO driven by monochromatic field 
 has been solved analytically up to now  in terms of the Fokker-Planck equation
in the complex-P representation and in the
steady-state regime by consideration of all orders of dissipation and decoherence.  Applications of these results to some oscillatory models  are given in Refs. \cite{Drummond}, \cite{a33}.

We shall determine the excitation number and the Wigner functions of oscillatory mode numerically on the base of master equation by means of the quantum state diffusion method \cite{qsd}. For the semiclassical approach, we shall calculate the Lyapunov exponent and the Poincar\'e section according to the framework of Eq. (\ref{semclass}). For the Lyapunov exponent, the analysis is performed on the semiclassical time series determined from Eq. (\ref{semclass}) according to \cite{sprott}, which gives the maximum Lyapunov exponent. Specifically, the definition of the Lyapunov exponent is given by $L=\frac{1}{\Delta t}\ln\frac{\|x_2(t)-x_1(t)\|}{\|x_2(t_0)-x_1(t_0)\|}$, where $x=\left(Re(\alpha),Im(\alpha),\beta\right)$, with $\beta$ being the time variable defined through $d\beta/dt=1$ which augments Eq. (\ref{semclass}) to create an autonomous system. Note that $x_2$ and $x_1$ represent two trajectories that are very close together at the initial time $t_0$. Furthermore, $\Delta t = t - t_0$, with $t \rightarrow \infty$. For $L>0$ the system shows chaotic dynamics. $L=0$ corresponds to the case of conservative regular systems, and $L<0$ indicates that the dissipative system is regular. Since our system is dissipative, we would not expect $L=0$.

\subsection{NDO under time-modulated force}

\begin{figure}
\includegraphics[width=\textwidth]{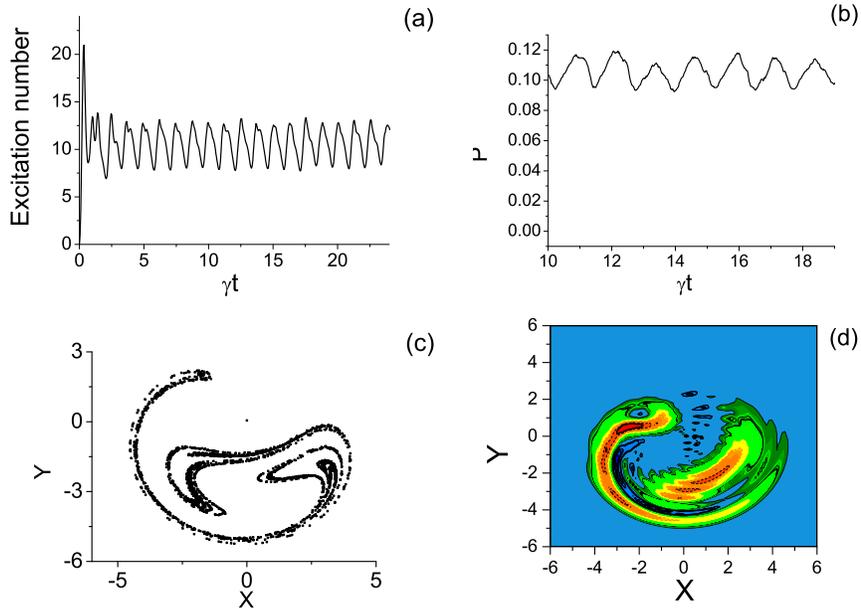} 
\caption{(a) The excitation number, (b) purity dynamics, and (c) Poincar\'e section, (d) Wigner function contour plot for the parameters:
$\Delta/\gamma=-15$, $\chi/\gamma=2$, $f_0/\gamma=10.2$,
$f_1/\gamma=10.2$, $\delta/\gamma=5$.}
\label{fig.sc_1}
\end{figure}

\begin{figure}
\includegraphics[width=\textwidth]{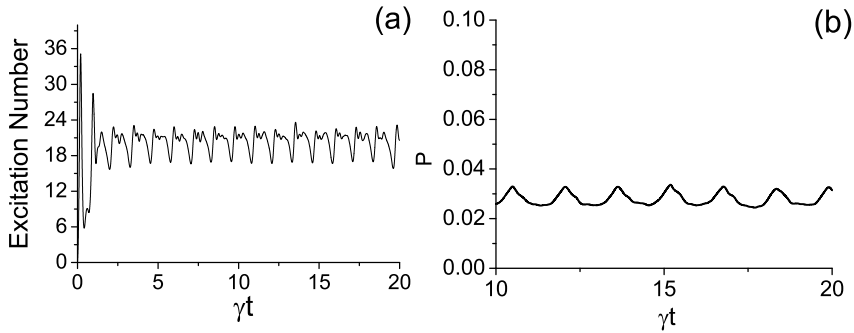}
\caption{(a) The excitation number, and (b) purity dynamics. The parameters are:
$\Delta/\gamma=-14.25$, $\chi/\gamma=0.175$, $f_0/\gamma=20.4$,
$f_1/\gamma=20.4, \delta/\gamma = 5$.}
\label{fig.purity_dynamics}
\end{figure}

Let us begin our study by considering a NDO under the action of a periodically modulated driving force, $f(t)=f_0+f_1\exp(\delta t)$. As shown in our earlier analysis \cite{PhysE}, for extended time scale exceeding the damping rate, the asymptotic dynamics
of the system is regular in the limit of small and
large values of the modulation frequency in comparison with the decay rate, and also when one of the perturbation forces $f_0$ and $ f_1$ is much larger than the other. Furthermore, the dynamics of the system is chaotic if the parameters $f_0$ and $f_1$ are approximately equal to each other. In Figs. \ref{fig.sc_1} and \ref{fig.purity_dynamics}, the mean excitation number of the averaged quantum trajectories and the corresponding purity for the same parameters versus time interval are plotted for parameters corresponding to the chaotic regime. In order to confirm that the regime is indeed chaotic,
we plot the classical Poincar\'e section in Fig.  \ref{fig.sc_1}(c) which displays a strange attractor with fractal structure typical of chaotic dynamics. Note that to calculate the Poincar\'e section, we have chosen $x_0$ and $y_0$ as an arbitrary initial phase space point of the system at the time $t_0$. We then
define a constant phase map in the $(X, Y)$ plane by the sequence of points
$(X_n, Y_n)=(X(t_n), Y(t_n))$ at $t_n=t_0+\frac{2\pi}{\delta}n$ ($n=0, 1, 2,
...$). Demonstration of chaos in the quantum treatment is presented in Fig. \ref{fig.sc_1}(d) via the Wigner function. The Wigner function exhibits a helical structure with contour plots concentrating approximately around the attractor, reflecting the intrinsic presence of chaotic dynamics.

Next, we present results of the mean excitation number and the purity of two chaotic regimes in greater detail (see Figs. \ref{fig.sc_1} and \ref{fig.purity_dynamics}). From the figures, we observe a time-dependent modulation with a period of $2\pi/\delta$ which corresponds to the period of the driving force at the over-transient time interval. Comparing the results of the two figures, we conclude that the magnitude of the purity for the regime of NDO shown in Fig. \ref{fig.purity_dynamics}(b) is lower than that in  Fig. \ref{fig.sc_1}(b). This outcome probably reflects the fact that for the second case, the NDO is in a deeper chaotic regime characterized by a smaller purity even when the excitation number depicted in Fig. \ref{fig.purity_dynamics}(a) does not essentially deviate from that of Fig. \ref{fig.sc_1}(a).

\begin{figure}
\includegraphics[width=\textwidth]{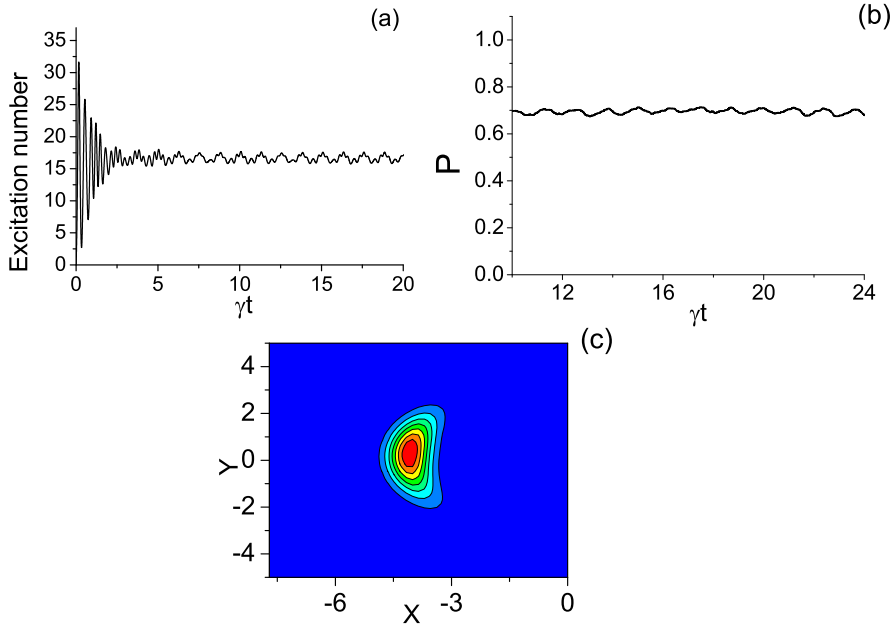}
\caption{(a) The excitation number, (b) purity dynamics, and (c) Wigner function contour plot for the parameters:
$\Delta/\gamma=-15$, $\chi/\gamma=0.7$, $f_0/\gamma=32.2$,
$f_1/\gamma=10.2$, $\delta/\gamma=5$.}
\label{fig.regular}
\end{figure}

In order to show that the lower purity is a result of chaotic dynamics, we present the purity dynamics for the case of regular dynamics in Fig.
\ref{fig.regular}(b). For the sake of comparison, we select the parameters for both the regular and chaotic regimes so that they have approximately the same oscillatory excitation number as shown in Figs. \ref{fig.sc_1}(a) and \ref{fig.regular}(a). Note that the relatively small amplitude of the modulation in the mean excitation number and the purity is due to the small ratio $f_1/f_0 $ chosen in this regime. 
The regularity of the dynamics is confirmed via calculation of the Wigner function (see Fig. \ref{fig.regular}(c)) which is almost Gaussian in this case. By comparing the results of Fig. \ref{fig.regular} with those of Figs. \ref{fig.sc_1} and \ref{fig.purity_dynamics}, we observe that the level of purity of the oscillatory mode for regular dynamics $P \approx 0.7$, essentially exceeds that of $P \approx 0.09$ and $P \approx 0.03$ for the case of chaotic dynamics.

\subsection{NDO driven by Gaussian  pulses}
First, we briefly dicuss the regimes of NDO driven by a sequence of Gaussian pulses. Depending on the parameters, we observe that this system is able to exhibit both regular and chaotic dynamics. The phenomenon of chaos is found to occur when the strength of the pulse trains $\Omega$ is sufficiently large, with the nonlinearity strength $\chi$ and pulse duration $T$ assuming appropriate values. In fact, this system was found to possess chaotic, deep chaotic and regular regime in an analogous fashion to the system discussed in the last section. However, in this section, we shall explore new regimes within this system, that will give rise to a low-level of excitation number as in the previous section. The other novelty in this section is the investigation on the connection between the purity of the quantum states and the Lyapunov exponent of the semiclassical dynamics. 

\begin{figure}
\includegraphics[width=\textwidth]{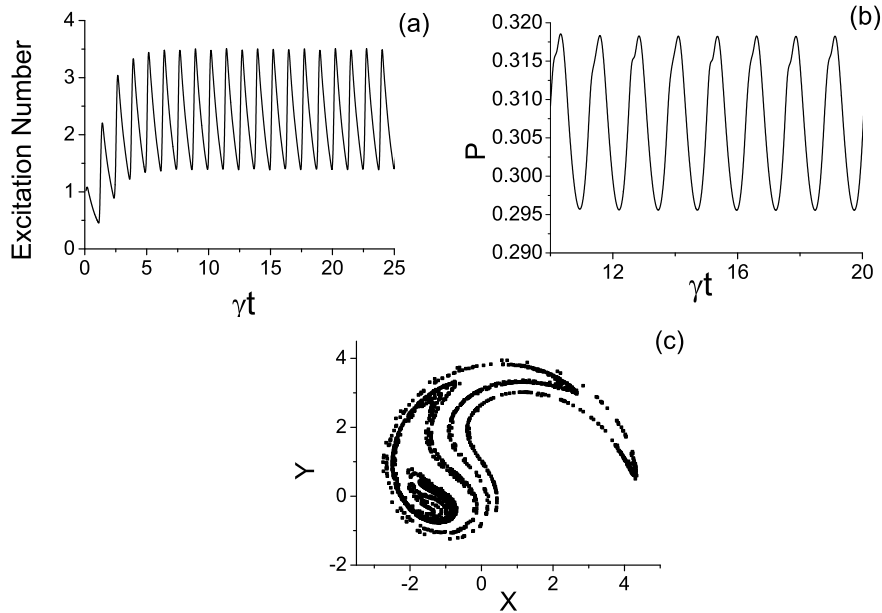} 
\caption{(a) The excitation number, (b) purity dynamics, and (c) Poincar\'e section, for the parameters:
$\Omega/\gamma=15$, $\chi/\gamma=0.7$, $\Delta/\gamma=-15$, $\gamma T=10.2$, $\gamma\tau=2\pi/5$.}
\label{fig.chaos_gaussianpulse}
\end{figure}

\begin{figure}
\includegraphics[width=\textwidth]{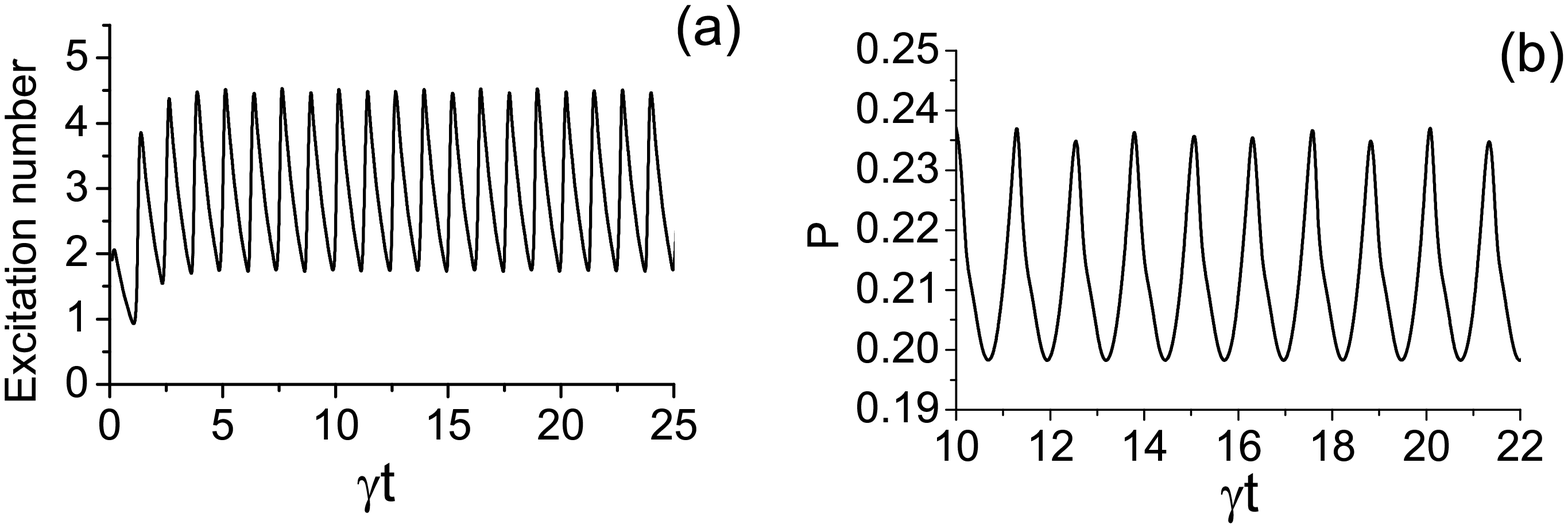} 
\caption{(a) The excitation number, and (b) purity dynamics. The parameters are:  $\Omega/\gamma=20.4$, $\chi/\gamma=0.7$, $\Delta/\gamma=-15$, $\gamma T=0.1$, $\gamma\tau=2\pi/5$.}
\label{fig.deepchaos_gaussianpulse}
\end{figure}

\begin{figure}
\includegraphics[width=\textwidth]{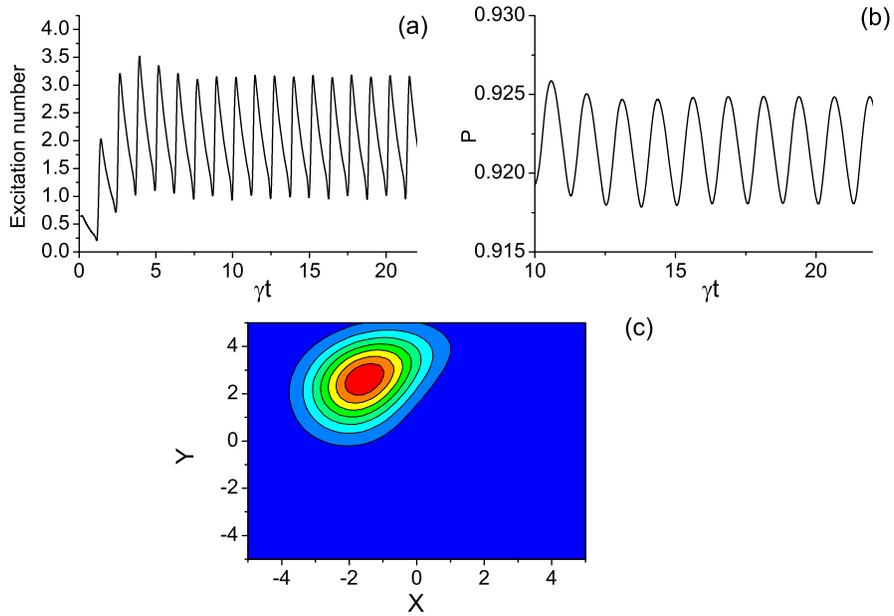} 
\caption{(a) The excitation number, (b) purity dynamics, and (c) the Wigner function indicating regular dynamics, for the parameters: $\Omega/\gamma=12$, $\chi/\gamma=0.1$, $\Delta/\gamma=-15$, $\gamma T=0.1$, $\gamma\tau=2\pi/5$.}
\label{fig.regular_gaussianpulse}
\end{figure}

As before, we shall adjust the parameters while keeping the oscillatory excitation number to be approximately constant. The results are shown in Figs. \ref{fig.chaos_gaussianpulse}, \ref{fig.deepchaos_gaussianpulse} and \ref{fig.regular_gaussianpulse}. As illustrated in Fig. \ref{fig.chaos_gaussianpulse}, the semiclassical dynamics display a strange chaotic attractor in the Poincar\'e section, with a Lyapunov exponent of $0.187$. Note that the excitation number and purity take an average value of about $2.5$ and $0.305$ respectively in this case. Then, by setting the system into the deep chaotic regime (with a Lyapunov exponent of $0.4197$), we observe a drop in the purity to a mean value of $0.22$ with an average excitation number of about $3$. This reduction in the purity as the excitation number is about the same and the corresponding semiclassical dynamics become more chaotic is consistent with the results obtained for time-periodic modulated driving discussed in the last section. And when we put the system in the regular domain (Lyapunov exponent $= -0.1693$) while maintaining the mean excitation number at $2$, we notice a sharp increase in the purity to an average value of $0.922$. Thus, we have again observed a larger purity for regular dynamics against chaotic dynamics as we fix oscillatory excitation number within a close range, which indicates a direct relationship between the purity of the state and the underlying oscillatory dynamics.

\begin{figure}
\includegraphics[width=8.6cm]{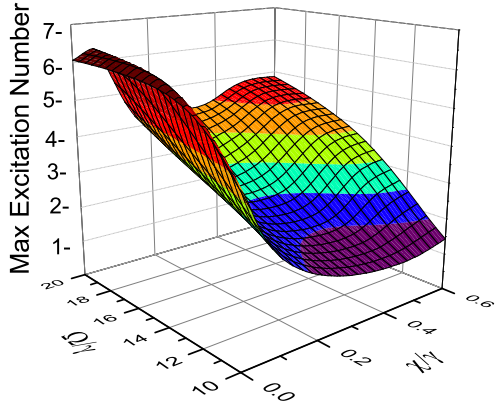} 
\caption{A plot of the maximum oscillatory excitation number against the parameter $\Omega/\gamma$ and $\chi/\gamma$. The values of the rest of the parameters are:$\Delta/\gamma=-15$, $\gamma T=0.1$, $\gamma\tau=2\pi/5$.}
\label{fig.MaxExNum_OmChi}
\end{figure}

\begin{figure}
\includegraphics[width=8.6cm]{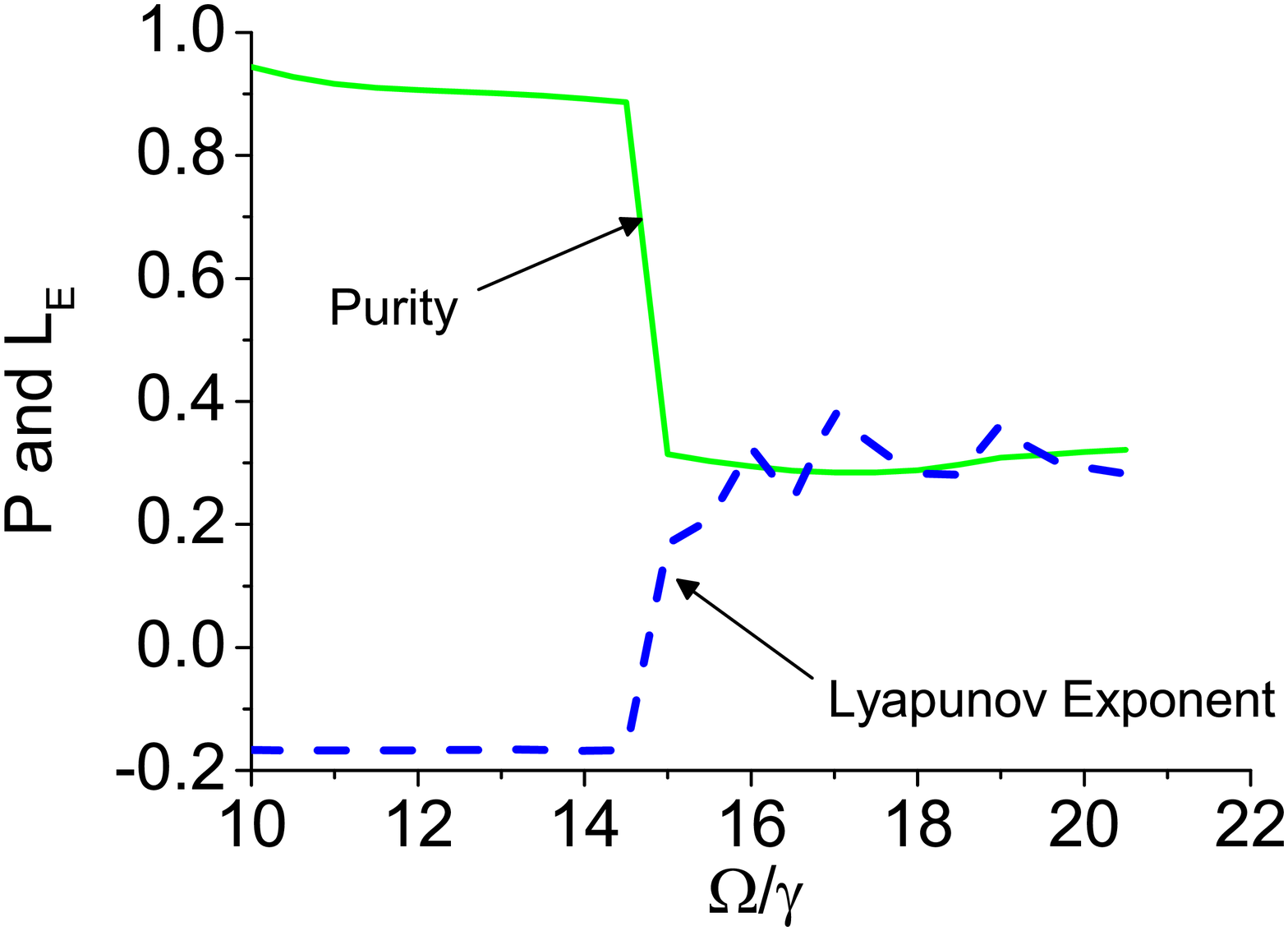} 
\caption{The maximum purity $P$ of the quantum states and the associated Lyapunov exponents of the corresponding semiclassical dynamical behaviour, versus the strength of the Gaussian pulse train $\Omega/\gamma$. Note that for each $\Omega/\gamma$, we have selected a specific $\chi/\gamma$ to ensure that the maximum excitation number lies within the range $3.6958$ to $5.5217$. The values of the rest of the parameters are: $\Delta/\gamma=-15$, $\gamma T=0.1$, $\gamma\tau=2\pi/5$.}
\label{fig.LyaExpPurity_OmChi}
\end{figure}

Let us now examine this relationship in greater detail. To do this, we first obtain a plot of the maximum oscillatory excitation number against the parameters $\Omega/\gamma$ and $\chi/\gamma$, which is shown in Fig. \ref{fig.MaxExNum_OmChi}. This allows us to determine the set of $\Omega/\gamma$ and $\chi/\gamma$ that gives approximately the same maximum excitation number. We take the set of $\Omega/\gamma$ and $\chi/\gamma$ that gives excitation number in the range $3.6958$ to $5.5217$. With this, we plot the maximum purity and Lyapunov exponent against $\Omega/\gamma$, which is illustrated in Fig. \ref{fig.LyaExpPurity_OmChi}. We observe the subtle relationship between purity and the corresponding semiclassical dynamical behaviour in this figure. When the semiclassical dynamics is regular, the purity is high; and when the semiclassical dynamics is chaotic, the purity reduces to a low value. In other words, there is a close link between an enhanced mixedness in the quantum domain with the presence of dynamical chaos in the semiclassical domain. In Fig. \ref{fig.LyaExpPurity_OmChi}, we observe the occurrence of a sharp transition, which happens at the same strength of the Gaussian driving field for both the purity and the Lyapunov exponent. These results illustrate the good quantum-to-classical correspondence in this system.

\section{Summary}
We have applied the purity of an ensemble of oscillatory states to analyse and identify the chaotic behaviour of two different NDO systems. In particular, we have calculated varying levels of purity of quantum states for both the regular and chaotic regimes of NDO driven by a continuously
modulated pump field, as well as for NDO under the action of a periodic sequence of identical pulses with Gaussian envelopes. It is important to note that these models seem feasible experimentally, and they can be achieved in several practial schemes involving nano-mechanical
oscillators as well as superconducting devices based on nonlinearity of the Josephson junction. In fact, we believe that
the NDO scheme operating with trains of Gaussian pulses is closer to practical realization via laser- or microwave pulses
(see, for example \cite{claud}  and \cite{xia}).

Finally, we have demonstrated that for both systems, the level of purity of oscillatory modes for regular dynamics essentially exceeds the analogous case of chaotic dynamics. Moreover, we have shown the occurrence of a drop in purity when the NDO moves into the deep chaotic regime which is characterized by a larger positive Lyapunov exponent.


\begin{thebibliography}{0}

\bibitem {Haake} F. Haake, Quantum Signatures of Chaos, Springer (2001).
\bibitem {Braun} D. Braun, Quantum Chaos and Decoherence, Springer (2001).
\bibitem {Cvitanovich }P.Cvitanovic’, R. Artuso, R. Mainieri, G. Tanner and G. Vattay, Chaos: Classical and Quantum, Chaos, Niels Bohr Institute, Copenhagen (2008).
\bibitem {Graham} E. Ott, M. Antonsen, Jr., and J. P. Hanson, Phys. Rev. Lett.
53, 2187 (1984); T. Dittrich and R. Graham, Phys. Rev. A 42, 4647 (1990).
\bibitem {Zurek} W. H. Zurek and J. P. Paz, Phys. Rev. Lett. 72, 2508 (1994);
75, 351 (1995); S. Habib, K. Shizume, and W. H. Zurek, Phys. Rev. Lett.
80, 4361 (1998); J. Gong and P. Brumer, Phys. Rev. E 60,
1643 (1999)
 \bibitem {PhysE} H. H. Adamyan, S. B. Manvelyan and G. Yu. Kryuchkyan, Phys. Rev. E, 64, 046219, (2001).

\bibitem {mpop} T. V. Gevorgyan, A. R. Shahinyan, G. Yu. Kryuchkyan, Dissipative Chaos in Quantum Distributions, In ”Modern
Optics and Photonics: Atoms and Structured Media. pp.60-77,  Eds: G. Kryuchkyan, G.
Gurzadyan and A. Papoyan, World Scientific, 2010.

\bibitem {Leon} A. Kowalewska-Kudłaszyk, J.K. Kalaga, W. Leonski, Phys. Rev. E  78,  066219 (2009).

\bibitem {K2} G. Yu. Kryuchkyan and S. B. Manvelyan, Phys. Rev. Lett., 88, 094101 (2002).
\bibitem {K3} G. Yu. Kryuchkyan and S. B. Manvelyan, Phys. Rev. A, 68, 013823 (2003).

\bibitem {Lloyd} J. Emerson, Y.S. Weinstein, S. Lloyd and D.G. Cory. Phys.Rev.Lett., 89, 284102-1, 2002;
 Y.S. Weinstein S. Lloyd and C. Tsallis. Phys.Rev.Lett., 89, 214101-1, 2002; A. Kowalewska-Kudłaszyk, J.K. Kalaga, W. Leonski, Phys. Lett., A 373, 1334 (2009).

\bibitem {Kulback}  A. Kowalewska-Kud laszyk, J. K. Kalaga, W. Leo´nski and V. Cao Long,    Phys. Lett.,  A376, 1280 (2012).

\bibitem{AMK} H. H. Adamyan, S. B. Manvelyan and G. Yu. Kryuchkyan, Phys. Rev. A, 63, 022102, (2001).

\bibitem {qsch} T. V. Gevorgyan, A. R. Shahinyan, and G. Yu. Kryuchkyan, Phys. Rev. A, 79, 053828, (2009).

\bibitem {Fock} T. V. Gevorgyan, A. R. Shahinyan, and G. Yu. Kryuchkyan, Phys. Rev. A, 85, 053802, (2012).

\bibitem{craig} H. G. Craighead. Science, 290, 1532 (2000).

\bibitem{ekinci} K. L. Ekinci and M. L. Roukes. Rev. of Scientific Instruments,  76, 061101 (2005).
\bibitem{claud} J. Claudon, et al., Phys. Rev. B 78, 184503 (2008).

\bibitem{vijay} R. Vijay, M. H. Devoret, and I. Siddiqi, Phys. Rev. A 80, 111101 (2009).

\bibitem{Nori} J. Q. You and F. Nori, Nature 474, 585 (2011).

\bibitem {Drummond} P. D. Drummond and D. F. Walls, J. Phys. A: Math. Gen {\bf 13}, 725 (1980).

\bibitem {a33} G. Yu. Kryuchkyan and K. V. Kheruntsyan, Opt.
Comm. {\bf 120}, 132 (1996); K. V. Kheruntsyan, D.S. Krahmer, G.Yu. Kryuchkyan, K.G. Petrossian, Opt.
Comm. {\bf 139}, 157 (1997); G. Yu. Kryuchkyan, K. V.  Kheruntsyan, V. 0. Papanyan and K. G. Petrossian, Quantum Semiclass. Opt. {\bf 7}, 965 (1995). 

\bibitem {qsd} I. C. Percival, Quantum State Diffusion(Cambridge
University Press, Cambridge), (2000).

\bibitem{sprott} J. C. Sprott, Chaos and Time-Series Analysis, Oxford University Press, Oxford, (2003).

\bibitem {xia} Xiaoting Wang, S. Vinjanampathy, F. W. Strauch, and
K. Jacobs, Phys. Rev. Lett. 107, 177204 (2011).



\end{thebibliography}
\end{document}